\documentclass{osa-article}
\usepackage{hyperref}
\usepackage{rotating}

\journal{oe}
\articletype{Research Article}
\begin{document}

\title{Parametric behavior of diffraction on guided-mode resonant grating of subwavelength thickness}

\author{Efremova E.,\authormark{1,2} Perminov S.,\authormark{3} and Vergeles S.S.\authormark{4,5,*}}

\address{
\authormark{1}Saint Petersburg State University, 7/9 Universitetskaya nab., St. Petersburg, 199034 Russia\\
\authormark{2}ITMO University, 49 Kronverksky Pr., St. Petersburg 197101, Russia\\
\authormark{3}A.V. Rzhanov Institute of Semiconductor Physics, Siberian Branch, Russian Academy of Sciences, 13 Lavrent’yev Avenue, Novosibirsk 630090, Russia\\
\authormark{4}Moscow Institute of Physics and Technology, Dolgoprudnyj, Institutskij lane 9, Moscow
Region 141700, Russia\\
\authormark{5}Landau Institute for Theoretical Physics, Russian Academy of Sciences, 1-A Akademika Semenova av., 142432 Chernogolovka, Russia}

\email{\authormark{*}ssver@itp.ac.ru} %% email address is required



\begin{abstract}
We consider Wood anomalies in diffraction spectrum from two-dimensional dielectric periodic grid embedded in a surrounding media. The grid is of subwavelength thickness, and diffraction of wave having S-polarization is investigated in the vicinity of the emergence of first diffraction maximum. We reduce Maxwell equations to coupled-mode theory with three parameters, which are integral characteristics of material arrangement in grid. In particular, we show that such grids are capable to have full reflectance in parametrically narrow frequency bandwidth. The effect is accompanied by a parametric evanescent field enhancement in the region near the grid. In particular, we consider grids with sinusoidal profile and show that this type of grids possesses unique diffraction properties due to absence of the guided mode coupling. For such grids, there is a thin transparency window at a background of near to zero transmission at slightly nonnormal incidence. We estimate what are enough grid size and the incident beam flatness to resolve the singularities.
\end{abstract}

%%%%%%%%%%%%%%%%%%%%%%%%%%  body  %%%%%%%%%%%%%%%%%%%%%%%%%%
\section{Introduction}

Wood anomalies, which stem from resonance with surface modes of guided mode resonance grating \cite{hessel1965new}, have numerous applications in optics \cite{loewen1997diffraction}. %\cite{darweesh2018role}
Here we consider two-dimensional dielectric periodic grid embedded in a surrounding media. The grid is of subwavelength thickness, and diffraction of wave having S-polarization is investigated near the emergence of first diffraction maximum. We show that the exact Maxwell equations can be subsequently reduced to coupled-mode theory with three parameters, which are integral characteristics of material arrangement in grid. In particular, we show that such a grids are capable to have a full reflectance in parametrically narrow frequency bandwidth. The effect accompanied by a parametric evanescent field enhancement in the region near the grid. We check our analytical predictions by a direct solution of the Maxwell equations.

Our analytical scheme has much in common with those applied in \cite{bykov2015spatiotemporal,bykov2017coupled}. The difference is all governing parameters are obtained directly from dielectric permittivity distribution in our case rather than to be phenomenologically determined in \cite{bykov2015spatiotemporal,bykov2017coupled}. Besides, we consider grids with sinusoidal profiles and show that this type of grids possesses unique diffraction properties due to absence of guided mode coupling. We show that there is a thin transparency window at a background of near to zero transmission at nonnormal incidence.

To resolve the mentioned singularities, one should use the grid of enough size and the beam with the enough wavefront flatness. We make corresponding estimates. Due to the system is effectively two-dimensional, it is possible to check the estimations by a solution of Maxwell equations in dipole-dipole approximation for a grid of several hundreds of the lattice periods.

\section{General relations}

We assume that the plate is located near surface $z=0$ and has periodical structure with period $L$ in $x$-direction. Its structure is described by dielectric permittivity $\varepsilon(x,z)=\varepsilon(x+L,z)$. The plate is uniform in $y$-direction. The thickness of the plate is $a$, thus the dielectric permittivity $\varepsilon=\varepsilon_{out}>0$ at $|z|>a/2$. We consider the limit of thin plates, when the wavelength inside the plate is small as compared to its thickness, $\sqrt{\varepsilon_{out}(\varepsilon_0+1)}\omega a/c\ll1$, where $\varepsilon_{out}(\varepsilon_0+1)$ is spaced-averaged value of $\varepsilon$ inside the plate at $|z|<a/2$, $\omega$ is the frequency of the light and $c$ is the speed of light in vacuum.

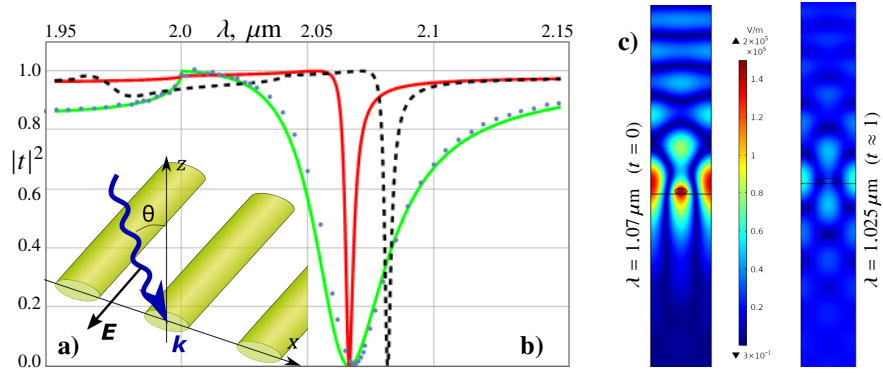
\begin{figure}[t]
\centering\noindent
\begin{picture}(300,135)
    \put(220,0){\begin{picture}(100,135)
        \put(8,0){\includegraphics[width=0.9cm]{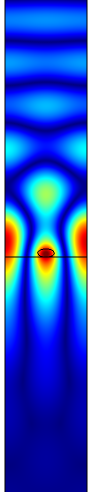}}
        \put(0,30){\rotatebox{90}{{\scriptsize $\lambda = 1.07\,\mu\text{m}$\ \ \  ($t=0$)}}}
        \put(65,0){\includegraphics[width=0.8cm]{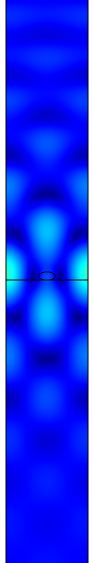}}
        \put(90,30){\rotatebox{90}{{\scriptsize $\lambda = 1.025\,\mu\text{m}$\ \ \  ($t\approx 1$)}}}
        \put(40,0){\includegraphics[width=0.5cm]{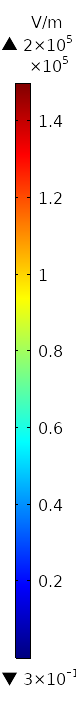}}
        \end{picture}}
    \put(0,0){\begin{picture}(200,120)
        \put(0,0){\includegraphics[width=7cm]{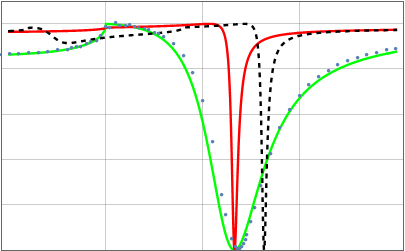}}
        \put(-10,0){\scriptsize{0.0}}
        \put(-10,22){\scriptsize{0.2}}
        \put(-10,44){\scriptsize{0.4}}
        \put(-10,63){\scriptsize{0.6}}
        \put(-10,90){\scriptsize{0.8}}
        \put(-10,110){\scriptsize{1.0}}
        \put(-13,75){$|t|^2$}
        \put(0,125){\scriptsize{1.95}}
        \put(47,125){\scriptsize{2.0}}
        \put(65,125){$\lambda, \,\mu\text{m}$}
        \put(97,125){\scriptsize{2.05}}
        \put(141,125){\scriptsize{2.1}}
        \put(188,125){\scriptsize{2.15}}
        \put(0,0){\includegraphics[width=3.5cm]{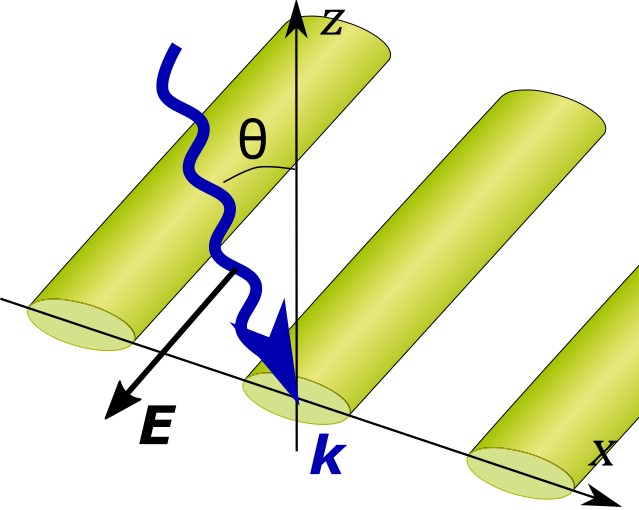}}
        \end{picture}}
    \put(217,120){\textbf{c)}}
    \put(5,5){\textbf{a)}}
    \put(180,5){\textbf{b)}}
\end{picture}
\caption{a) Scheme of the grid. The chosen parameters are period $L=2\,\text{\textmu m}$, the cross section of each rod is ellipse with semiaxes $0.2\,\text{\textmu m}$ and $0.1\,\text{\textmu m}$, thus $\Lambda a\varepsilon_0\approx 0.30$,  $\Lambda a\varepsilon_1\approx0.28$,  $\Lambda a\varepsilon_2\approx0.24$. Dielectric permittivity of the rod material is $\varepsilon=4\varepsilon_{out}$. b) Transition coefficient $|t|^2$ as a function of wavelength $\lambda$ in the surrounding medium. Green line corresponds to normal incidence, dots are the result of DDA with $N=500$ rods. Red curve corresponds to normal incidence, but $\varepsilon_1$ is diminished in 3 times with the same $\varepsilon_{0,2}$; dashed line is for $\theta=0.02$ rad. c) The field enhancement factor is $6.5$ at resonance wavelength $\lambda=2.07\,\mu\text{m}$, the region thickness in $z$-direction is $1/\kappa_1\approx2.7\,\mu\text{m}$}
\label{fig:one}
\end{figure}

In the paper, we consider S- or TE-polarization, when the electric field of the incident electromagnetic wave is directed along $y$-direction, see Fig.\ref{fig:one}a. Then the electric field has only nonzero $y$-component in all space due to the dielectric permittivity does not alter in $y$-direction, see Fig.1a). Let us decompose full electric field ${\bf E}$ onto incident wave ${\bf E}_{in}$ and scattered field ${\bf E}_{sc}$, ${\bf E}= {\bf E}_{in} + {\bf E}_{sc}$. First we assume the simplest model case, when the incident wave is plane wave normalized to unity, $E_{in} = \exp(-ikz\cos\theta + ikx\sin\theta)$, where $k=\sqrt{\varepsilon_{out}}\omega/c$ is the wavenumber in surrounding media and $\theta$ is the incidence angle. Here and below the scalar fields are $y$-components of the vector fields.  In our case the wave equation has the form of Schr\"{o}dinger equation with source,
\begin{equation}\label{waveeq}
    \left(
        (\partial_x^2 +\partial_z^2) + k^2 + \delta \varepsilon \cdot k^2
        \right)E_{sc}
    =
    \delta \varepsilon \cdot k^2E_{in},
    \qquad
    \delta \varepsilon = \varepsilon/\varepsilon_{out}-1.
\end{equation}

Consider first uniform plate, when $\varepsilon = \varepsilon_0$ everywhere at $|z|<a/2$. If the plate is a dielectric which is more optically dense than the medium, $\varepsilon_0>\varepsilon_{out}$, then there exists a guided mode which is bounded in $z$-direction. To find the spatial structure of the mode, one should solve (\ref{waveeq}) with zero r.h.s. Due to the plate is thin, one can replace exact permittivity distribution with approximate
\begin{equation}\label{deltauniform}
    \delta \varepsilon =
    \varepsilon_0a\delta(z)
\end{equation}
The replacement can be treated as a quantum mechanical $\delta$-potential, which mimicries the shallow one-dimensional quantum well just if $\sqrt{\varepsilon_0} ka\ll1$. The bounded eigen-function for this potential is $E_{sc} = \exp(-\kappa |z| + iqx)$ with $\kappa^2 = q^2-k^2$ and the dispersion relation
\begin{equation}\label{f}
    f = 0, \qquad f = 2\kappa/k - \varepsilon_0ka.
\end{equation}
The spatial profile of $E_{sc}$ can be approximated as constant in $z$-direction. Note that $\varepsilon_{out}\varepsilon_0a/4\pi$ can be treated as surface polarizability of the plate in external uniform electric field parallel to the plate.

We turn to periodically structured plate. In the same approximation (\ref{deltauniform}) one can represent the spatial distribution of the dielectric permittivity as (see analogous expansion \cite{benabbas2005analytical}, where the metal-dielectric structure has finite thickness)
\begin{equation}\label{deltavarepsilon}
    \delta \varepsilon =
    \Big(
        \varepsilon_0
        +
        \varepsilon_{1+} e^{i\Lambda x}
        +
        \varepsilon_{1-} e^{-i\Lambda x}
        +
        \varepsilon_{2+} e^{2i\Lambda x}
        +
        \varepsilon_{2-} e^{-2i\Lambda x}
        +
        \ldots
        \Big)a\delta(z),
\end{equation}
where the inverse lattice constant $\Lambda=2\pi/L$. Dimensionless quantities $\varepsilon_{i\pm}$ are generally complex. If the material remains lossy without gain, it should be fulfilled $\Im(\varepsilon)\geq 0$ in each point. We keep only first and second harmonics that is sufficient as it will be shown below. Now one should solve equation (\ref{waveeq}) with (\ref{deltavarepsilon}). It is natural to represent the scattered electric field as a discrete Fourier series
\begin{equation}
    E_{sc}
    =
    \exp(ipx)\sum\limits_{l=-\infty}^{+\infty}E_{sc,l}\exp(-\kappa_l|z| + il\Lambda x),
    \qquad p = k\sin\theta,
    \qquad
    \kappa_l = \sqrt{(l\Lambda+p)^2-k^2}
\end{equation}
Imaginary part of $\kappa_l$ should be nonpositive. Coefficients $E_{sc,l}$ satisfy the series of equations
\begin{equation}
    f_l E_{sc,l}
    -ka\sum \varepsilon_{i,\pm}E_{sc,l\mp i}
    =
    ka\varepsilon_0 E_{in,l} + ka\sum \varepsilon_{i,\pm}E_{in,l\mp i},
\end{equation}
and one should account that the only nonzero coefficient for the incident field is $E_{in,0} = 1$. Note that if $\kappa_l$ is imaginary, then $E_{sc,l}$ is the scattered wave amplitude running from the plate. In particular, $r=E_{sc,0}$ is the complex amplitude of refraction and $t=1+E_{sc,0}$ is the amplitude of transmission.

Now we assume that the plate is thin in a sense that $\varepsilon_{i\pm}ka\ll1$ as well. This restriction can be more strong for optically dense materials, if the $\varepsilon$ variance is of the order of its mean value. Under the assumption, the qualitative picture of $E_{sc,l}$ amplitude distribution is as follows. Since the frequency is close to emergence point of $\pm1$st diffraction maxima, typical values are $f_0 \approx -2i$, $f_{\pm1}\ll1$ and $f_{l}>1$ at $|l|\geq2$. This means that one can neglect all harmonics with $l\geq2$ in the main approximation and consider only three with $l=0,\pm1$. The corresponding coefficients satisfy the system of equations
\begin{eqnarray}\label{sys1}
    f_0 E_{sc,0} - ka\varepsilon_{1-}E_{sc,1}-ka\varepsilon_{1+}E_{sc,-1}
    =
    ka\varepsilon_0,
    \\ \label{sys2}
    f_{\pm1} E_{sc,\pm1} -ka\varepsilon_{1\pm}E_{sc,0} - ka\varepsilon_{2\pm}E_{sc,\mp1}
    =
    ka\varepsilon_{1\pm}.
\end{eqnarray}
compare with full infinite system of equations \cite{hessel1965new}. Here a slab surrounded by a media from both sides is considered, so we have deal with surface dipole susceptibility $\varepsilon_0 a$ and its off-diagonal matrix elements $\varepsilon_{i\pm}a$ instead of surface impedances $Z^s_i$. The solution of the system (\ref{sys1},\ref{sys2}) is for zeroth diffraction maximum
\begin{equation}\label{rF}
    r = \frac{\varepsilon_0 F + \varepsilon_{1+}\varepsilon_{1-}}{(f_0/ka)F-\varepsilon_{1+}\varepsilon_{1-}},
    \qquad
    F = \frac{f_1f_{-1}/(ka)^2 - \varepsilon_{2-}\varepsilon_{2+}}
    {(f_{1}+f_{-1})/ka +
        (\varepsilon_{1-}/\varepsilon_{1+})\varepsilon_{2+}
        + (\varepsilon_{1+}/\varepsilon_{1-})\varepsilon_{2-}}
\end{equation}
and
\begin{equation}
    E_{sc,\pm1}
    =
    \frac{(f_{\mp1}/ka)\varepsilon_{1\pm} + \varepsilon_{2\pm}\varepsilon_{1\mp}}
        {f_1f_{-1}/(ka)^2 - \varepsilon_{2-}\varepsilon_{2+}}
    t
\end{equation}
for $\pm1$st diffraction maxima. Developed scheme is similar to that used in \cite{bykov2015spatiotemporal,bykov2017coupled}. The difference is here we find explicitly the coefficients $\varepsilon_{i\pm}$ from the space distribution $\varepsilon(x)$.

\section{Results}

For the calculations with the finite grids we used the Discrete Dipoles Approximation (DDA) method. Originally, it was used in astrophysics to study interstellar dust \cite{Purcell73,Draine88} and since that this
technique is being used for a long time to study light scattering and near field distribution in nanophotonics. A comprehensive review and classification was done in \cite{Yurkin07}, while the details of derivation
and application of DDA in 2D case can be found in \cite{Martin98,praBNPFS17}. For the periodic system, DDA method was used in the recent study of the Wood-Rayleigh anomalies \cite{frumin2013plasmons}.

\begin{figure}[h!]
\centering
\begin{picture}(300,130)
\put(-35,0){\begin{picture}(150,130)
    \put(0,0){\includegraphics[width=6.5cm]{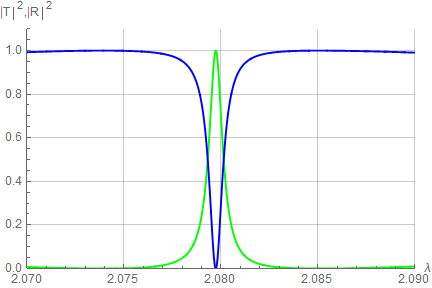}}
    \put(100,20){\includegraphics[width=3cm]{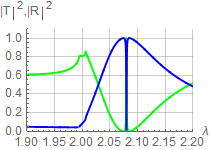}}
    \end{picture}}
\put(155,0){\begin{picture}(150,130)
    \put(0,0){\includegraphics[width=6.5cm]{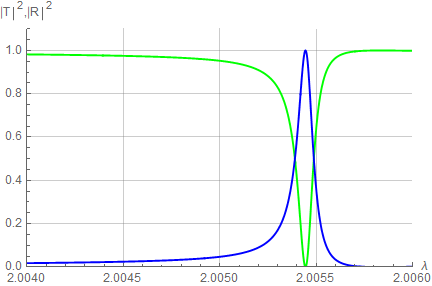}}
    \put(20,30){\includegraphics[width=3cm]{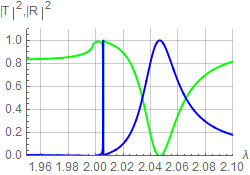}}
    \end{picture}}
\end{picture}
\caption{Fine structure of the transmission-reflection curves. a) Pure sinusoidal grating with $\Lambda a\varepsilon_0\approx 0.59$,  $\Lambda a\varepsilon_1\approx0.56$,  $\varepsilon_2=0$, the transparency window at $\theta=0.003$. b) The second blue-shifted resonance for grid with $\Lambda a\varepsilon_0\approx 0.3$, $\Lambda a\varepsilon_1\approx0.28$,  $\varepsilon_2=0.15$, at angle $\theta=10^{-3}$.}
\label{fig:two}
\end{figure}

Now we consider plate which is made of transparent dielectric. First suppose that the grating is pure sinusoidal, such that $\varepsilon_{i,\pm}=0$ for $i\geq2$. Such a grid which is thick slab was considered in e.g. \cite{chuang1983wave} and can be fabricated with interferometric lithography methods \cite{xia2011nanostructures}. Then one can choose the onset of $x$-axis in that way so $\varepsilon_{1+} = \varepsilon_{1-} \equiv \varepsilon_1$ is real, that is $\varepsilon(-x)=\varepsilon(x)$. We arrive to brief equation $ka/F= 1/f_1 + 1/f_{-1}$ for $F$ in (\ref{rF}) and
\begin{equation}\label{tdielectric}
    t = \Big(1-(i\varepsilon_1^2/2)\big(f_1^{-1} + f_{-1}^{-1}\big)\Big)^{-1},
    \qquad
    E_{sc,\pm1} = \frac{ka\varepsilon_1}{f_{\pm1}} t.
\end{equation}
Transition amplitude has a drop to zero when resonance (\ref{f}) is achieved in $l=1$ or $l=-1$ harmonics \cite{liu1998high}. In the case of normal wave incidence $\theta=0$ the points coincide being determined by $f_1=0$. They are redshifted from the point of appearance of $\pm1$st diffraction maxima, its position can be evaluated as $k_1\vert_{\theta=0}=\Lambda(1-(ka\varepsilon_0)^2/8)$ if $ka\varepsilon_0\ll1$. The thickness of the drop as a function of frequency is $\delta k_1\sim \Lambda(ka)^3\varepsilon_0\varepsilon_1^2$, its dependence on $\varepsilon_1$ is demonstrated in Fig.~\ref{fig:one}b. When the incidence is not normal, then the drop is split on two drops which are located at $k_{\pm1}(\theta) =k_1\vert_{\theta=0} \mp \Lambda\theta$. The drops correspond to zeros of $F$ (\ref{rF}) and thus to refraction amplitude $r=-1$. The drops are separated by a peak up to unity at arbitrary small angle $\theta$, the thickness of the transparency window is $\sim\Lambda\theta$, see Fig.~\ref{fig:two}a). The peak is determined by the condition that the phases of $E_{sc,\pm1}$ are opposite that is provided by $f_1=-f_{-1}$. Its nature is analogous to that of  electromagnetically induced transparency, see review \cite{limonov2017fano}. In real world, the peak should be smoothed due to grid imperfections, its finite size and %\textcolor{blue}
{deviations of the incident wave from ideally plane wave.}

Now consider grid with nonzero $\varepsilon_{2\pm}$. The case corresponds to any composite grid made of alternating materials with different optical properties. We restrict our consideration to the simplest case of symmetric grid $\varepsilon(-x)=\varepsilon(x)$, then $\varepsilon_{2-}=\varepsilon_{2+}\equiv \varepsilon_2$, which is assumed to be of the order of $\varepsilon_0$. In the case of normal wave incidence one arrives to (\ref{tdielectric}) at $\theta=0$ with replacement $f_1 \to 2\kappa_1/k - ka(\varepsilon_0+\varepsilon_2)$. This shifts the position of zero point $k_1\vert_{\theta=0}$ of transmission amplitude $t$, $\delta k_1\vert_{\theta=0}= \Lambda(1- (ka(\varepsilon_0+ \varepsilon_2))^2/8)$ and correspondingly the thickness of the drop $\delta k_1$. The independence of the zero point on $\varepsilon_1$ is demonstrated in Fig.~\ref{fig:one}b. Note that the near field is parametrically large when $t=0$ due to $E_{sc,\pm1} = -i(\varepsilon_1+\varepsilon_2)/(2\Lambda a\varepsilon_1^2)$ at corresponding frequency, see Fig.\ref{fig:one}c) and \cite{wei2006electric}. Thus, the electric field is enhanced near the grid in $2|E_{sc,1}|$ times. The thickness in $z$-direction of the region where the field achieves its maximum value is $1/\kappa_1= L^2/(2\pi^2a\varepsilon_0)$, so the region extends far beyond the grid. The difference increases for nonzero $\theta$. The main drop slightly redshifts, $k_1 - k_1\vert_{\theta=0} \sim -\Lambda \theta^2/(ka\varepsilon_2)^2$ at small angles $\theta\lesssim (ka\varepsilon_2)^2$. At these angles, the drop can be interpreted as a Fano resonance between low quality mode $l=0$ and high quality symmetric mode combination $E_{sc,1} + E_{sc,-1}$. If angle $\theta$ is greater, then the shift is $-\Lambda \theta$ as before, and now the resonance mode is just $E_{sc,1}$. The second drop appears at $\delta k_{-1}\vert_{\theta=0}= \Lambda(1-(ka(\varepsilon_0-\varepsilon_2))^2/8)$ for small angles and it is no more down to zero when $|\theta|>(ka)^2(\varepsilon_0^2-\varepsilon_2^2)/(16\varepsilon_0^2)$. The form of the drop corresponds to Fano resonance with anti-symmetric mode $E_{sc,1} - E_{sc,-1}$.
%\textcolor{blue}
{For the small angles, the thickness of the resonance is of the order of $\Lambda \theta^2/\big((ka)^2(\varepsilon_0-\varepsilon_2)\varepsilon_2\big)$.}

\section{Discussion}

Field enhancement at the interface of slab grid is show to be possible to utilize for biosensing \cite{el2010sensitivity}. Here we show that the field enhancement can be increased via diminishing of coupling of the guided modes with the incident wave, in our term this is $\varepsilon_1$. The fee for the field enhancement is the narrowing of the resonance in frequency and angle domains. Thus diminishing $\varepsilon_1$, one should use beam with less spectral width and of more flatness and a grid of larger size.

\section*{Funding}
The work was supported by RScF grant No.17-79-20418.

%\section*{Acknowledgments}
%Acknowledgments, if included, should appear at the end of the document. The section title should not be numbered.

%%%%%%%%%% If using BibTeX:
\bibliography{wa}

\begin{thebibliography}{10}
\newcommand{\enquote}[1]{``#1''}

\bibitem{hessel1965new}
A.~Hessel and A.~Oliner, \enquote{A new theory of wood’s anomalies on optical
  gratings,} {\protect\JournalTitle{Applied optics}} \textbf{4}, 1275--1297
  (1965).

\bibitem{loewen1997diffraction}
E.~G. Loewen and E.~Popov, \emph{Diffraction gratings and applications} (Marcel
  Dekker, New York, 1997).

\bibitem{bykov2015spatiotemporal}
D.~A. Bykov and L.~L. Doskolovich, \enquote{Spatiotemporal coupled-mode theory
  of guided-mode resonant gratings,} {\protect\JournalTitle{Optics express}}
  \textbf{23}, 19234--19241 (2015).

\bibitem{bykov2017coupled}
D.~A. Bykov, L.~L. Doskolovich, and V.~A. Soifer, \enquote{Coupled-mode theory
  and fano resonances in guided-mode resonant gratings: the conical diffraction
  mounting,} {\protect\JournalTitle{Optics express}} \textbf{25}, 1151--1164
  (2017).

\bibitem{benabbas2005analytical}
A.~Benabbas, V.~Halt{\'e}, and J.-Y. Bigot, \enquote{Analytical model of the
  optical response of periodically structured metallic films,}
  {\protect\JournalTitle{Optics Express}} \textbf{13}, 8730--8745 (2005).

\bibitem{Purcell73}
E.~M. Purcell and C.~R. Pennipacker, \enquote{Scattering and absorption of
  light by nonspherical dielectric grains,} {\protect\JournalTitle{Astrophys.
  J.}} \textbf{186}, 705 (1973).

\bibitem{Draine88}
B.~T. Draine, \enquote{The discrete-dipole approximation and it's application
  to the interstellar graphite grains,} {\protect\JournalTitle{Astrophys. J.}}
  \textbf{333}, 848 (1988).

\bibitem{Yurkin07}
M.~Yurkin and A.~Hoekstra, \enquote{The discrete dipole approximation: An
  overview and recent developments,} {\protect\JournalTitle{Journal of
  Quantitative Spectroscopy and Radiative Transfer}} \textbf{106}, 558 -- 589
  (2007). <ce:title>IX Conference on Electromagnetic and Light Scattering by
  Non-Spherical Particles</ce:title>.

\bibitem{Martin98}
O.~J.~F. Martin and N.~B. Piller, \enquote{Electromagnetic scattering in
  polarizable backgrounds,} {\protect\JournalTitle{Phys. Rev. E}} \textbf{58},
  3909--3915 (1998).

\bibitem{praBNPFS17}
A.~S. Bereza, A.~V. Nemykin, S.~V. Perminov, L.~L. Frumin, and D.~A. Shapiro,
  \enquote{Light scattering by dielectric bodies in the born approximation,}
  {\protect\JournalTitle{Phys. Rev. A}} \textbf{95}, 063839 (2017).

\bibitem{frumin2013plasmons}
L.~Frumin, A.~Nemykin, S.~Perminov, and D.~Shapiro, \enquote{Plasmons excited
  by an evanescent wave in a periodic array of nanowires,}
  {\protect\JournalTitle{Journal of Optics}} \textbf{15}, 085002 (2013).

\bibitem{chuang1983wave}
S.~Chuang and J.~Kong, \enquote{Wave scattering and guidance by dielectric
  waveguides with periodic surfaces,} {\protect\JournalTitle{JOSA}}
  \textbf{73}, 669--679 (1983).

\bibitem{xia2011nanostructures}
D.~Xia, Z.~Ku, S.~Lee, and S.~Brueck, \enquote{Nanostructures and functional
  materials fabricated by interferometric lithography,}
  {\protect\JournalTitle{Advanced materials}} \textbf{23}, 147--179 (2011).

\bibitem{liu1998high}
Z.~Liu, S.~Tibuleac, D.~Shin, P.~Young, and R.~Magnusson,
  \enquote{High-efficiency guided-mode resonance filter,}
  {\protect\JournalTitle{Optics letters}} \textbf{23}, 1556--1558 (1998).

\bibitem{limonov2017fano}
M.~F. Limonov, M.~V. Rybin, A.~N. Poddubny, and Y.~S. Kivshar, \enquote{Fano
  resonances in photonics,} {\protect\JournalTitle{Nature Photonics}}
  \textbf{11}, 543 (2017).

\bibitem{wei2006electric}
C.~Wei, S.~Liu, D.~Deng, J.~Shen, J.~Shao, and Z.~Fan, \enquote{Electric field
  enhancement in guided-mode resonance filters,} {\protect\JournalTitle{Optics
  letters}} \textbf{31}, 1223--1225 (2006).

\bibitem{el2010sensitivity}
M.~El~Beheiry, V.~Liu, S.~Fan, and O.~Levi, \enquote{Sensitivity enhancement in
  photonic crystal slab biosensors,} {\protect\JournalTitle{Optics express}}
  \textbf{18}, 22702--22714 (2010).

\end{thebibliography}

%%%%%%%%%% If preparing manually:
% \newcommand{\enquote}[1]{``#1''}

% \bibitem{Zhang:14}
% Y.~Zhang, S.~Qiao, L.~Sun, Q.~W. Shi, W.~Huang, L.~Li, and Z.~Yang,
%   \enquote{Photoinduced active terahertz metamaterials with nanostructured
%   vanadium dioxide film deposited by sol-gel method,}
%   {\protect\JournalTitle{Optics Express}} \textbf{22}, 11070--11078 (2014).

% {Optical Society}, \enquote{{OSA Publishing},}
%   \url{http://www.osapublishing.org}.

% \bibitem{FORSTER2007}
% P.~Forster, V.~Ramaswamy, P.~Artaxo, T.~Bernsten, R.~Betts, D.~Fahey,
%   J.~Haywood, J.~Lean, D.~Lowe, G.~Myhre, J.~Nganga, R.~Prinn, G.~Raga,
%   M.~Schulz, and R.~V. Dorland, \enquote{Changes in atmospheric consituents and
%   in radiative forcing,} in \enquote{Climate Change 2007: The Physical Science
%   Basis. Contribution of Working Group 1 to the Fourth assesment report of
%   Intergovernmental Panel on Climate Change,}  S.~Solomon, D.~Qin, M.~Manning,
%   Z.~Chen, M.~Marquis, K.~B. Averyt, M.~Tignor, and H.~L. Miler, eds.
%   (Cambridge University Press, 2007).

% \end{thebibliography}

\end{document}